\documentclass[12pt]{article}

\usepackage{tabularx}
\usepackage{amsmath}
\usepackage[dvips]{epsfig}
\usepackage{amssymb}
\newcommand{\beq}[1]{\begin{equation}\label{#1}}
\numberwithin{equation}{section}

\newcommand{\reff}[1]{{\rm (\ref{#1})}}

\def\Proof{{\sl Proof:} }
\def\cj{e^{\o_j\a_j+4\o_j^3t}}

\def\intp{\int_t^\infty}
\def\vs{\vskip 2mm}

\def\C{{\mathbb C}}
\def\R{\mathbb R}

\def\fr{\frac}
\def\e{\varepsilon}
\def\a{\alpha}
\def\b{\beta}
\def\g{\gamma}
\def\d{\delta}

\def\l{\lambda}
\def\o{\omega}

\def\vphi{\varphi}
\def\wh{\widehat}
\def\s{\sigma}
\def\inmx{\int_{-\infty}^x}
\def\inmp{\int_{-\infty}^\infty}
\def\inxp{\int_x^{\infty}}

\def\AA{{\cal A}}
\def\BB{{\cal B}}

\def\RR{{\cal R}}

\def\sech{{\rm sech}}
\def\tanh{{\rm tanh}}
\def\WW{{\cal W}}
\def\KK{{\cal K}}

\def\sqr#1#2#3{{{\vbox{\hrule height.#2pt
            \hbox{\vrule width.#2pt height#1pt \kern#3pt
            \vrule width.#2pt}\hrule height.#2pt}}}}
\def\qed{$\ \ \mathchoice\sqr644\sqr644\sqr{2.1}3{2.1}\sqr{1.5}3{1.5}$}

\newtheorem{theorem}{Theorem}[section]
\newtheorem{lemma}[theorem]{Lemma}
\newtheorem{corollary}[theorem]{Corollary}

\newcommand{\blem}[1]{\begin{lemma}\label{#1}}
\newcommand{\elem}{\end{lemma}}
\newcommand{\btheo}[1]{\begin{theorem}\label{#1}}
\newcommand{\etheo}{\end{theorem}}
\newcommand{\bcor}[1]{\begin{corollary}\label{#1}}
\newcommand{\ecor}{\end{corollary}}

\newcommand{\lmat}{\begin{pmatrix}}
\newcommand{\rmat}{\end{pmatrix}}

\begin{document}

\title{Inversion of the linearized Korteweg-deVries equation 
at the multi-soliton solutions}

\author{M. H\u ar\u agu\c s-Courcelle\thanks{Supported by the Deutsche
Forschungsgemeinschaft  under
Ki 131/12-1 and Ki 131/12-2.}\\
Universit\"at Stuttgart\\70569 Stuttgart, Germany
\and
D. H. Sattinger\thanks{Research supported by the National Science
Foundation under Grant  DMS-9501233.}\\
University of Minnesota\\Minneapolis 55455, USA}

\date{\today}

\maketitle

{\abstract
Uniform estimates 
for the decay structure of the $n$-soliton 
solution of the Korteweg-deVries equation are obtained.  
The KdV equation, linearized at the $n$-soliton
solution is investigated in a class $\WW$ consisting of sums
of travelling waves plus an exponentially decaying residual term.
An analog of the kernel of the time-independent equation  
is proposed, 
leading to solvability conditions on the inhomogeneous term.
Estimates on the inversion of the linearized KdV equation at
the $n$-soliton are obtained. }


\vs
\centerline{Zeit. Angew. Math. Phys (ZAMP), {\bf 49}, (1998), 436-469}

\section{Introduction}\label{s1}

A little over 100 years ago Korteweg and deVries \cite{KdV} derived their now-famous equation
$$
u_t+u_{xxx}-6uu_x=0,
$$
in order to explain the observations of solitary waves in barge canals
by J. S. Russell \cite{Ru}.
Their equation exhibited solitary waves, and so
showed that such phenomena can
plausibly be explained by the Euler equations governing the motion of 
gravity
waves in an inviscid fluid.\footnote{Recently Pego and Weinstein
\cite{PW2} have
discovered that a simple equation equivalent to the KdV equation
appears in Boussinesq's treatise \cite{Bo}, and, moreover, that Boussinesq had
noted the existence of solitary waves. }
This not only resolved a long-outstanding
controversy, but
eventually led to fundamental new developments in mathematics.

In 1967 Gardner, Greene, Kruskal, and Miura \cite{GGKM} showed that the KdV
equation
is in fact a Hamiltonian equation possessing remarkable properties.
It could be solved by the method of inverse scattering, using the
scattering Schr\"odinger equation
$$
\psi_{xx}+k^2\psi-u(x,t)\psi=0.
$$
As $u(x,t)$ evolves according to the nonlinear KdV equation, the
scattering
data of the associated Schr\"odinger equation evolves linearly.

It is still an open problem as to the extent to which solutions of the KdV
equation approximate solutions of the full Euler equations. There is an extensive literature on  
the validity of the KdV approximation to solitary waves
of the full Euler equations; cf. for example
the works of Amick and Kirchg\"assner \cite{AK}, 
Kirchg\"assner \cite{Ki}; see also \cite{Be}, \cite{BL}.
However, in a moving reference
frame,
the solitary wave appears as a time independent solution, and is
considerably easier to treat mathematically than perturbations of
fully time dependent solutions of the KdV equation.

Craig \cite{Cr} has shown that the KdV equation is a good
approximation
to the full Euler equations over a time period of order
$\epsilon^{-3}$,
where $\epsilon^2=\lambda-1$, $\lambda$ being the inverse square of
the Froude number (see also Kano and Nishida \cite{KN}). 
It is highly unlikely, however, that the
KdV equation is an accurate approximation to the Euler equations over
an infinite time scale.

One natural question which one may pose is the following.
The KdV equation possesses the
so-called $n-$soliton
(multi-soliton) solution

\begin{equation}\label{n-sol}
u(x,t)=-2\fr{d^2}{dx^2}\log \det \Big( \d_{jk}+\fr{
e^{-(\theta_j+\theta_k)}}{\o_j+\o_k}\Big),
\qquad
\theta_j=\o_j(x-\a_j-4\o_j^2t),
\end{equation}
where $0<\o_1<\dots<\o_n$ and $\a_j\in\R$, $j=1,\dots,n$.

One may ask, do the multi-soliton solutions of the KdV equation extend to full,
stable
solutions of the Euler equations, say on a semi-infinite time axis
$0\le t\le \infty$? A positive answer to this question would, as a by product,
prove the existence of non-trivial time dependent solutions to the Euler
equations on a semi-infinite time interval $0<t<\infty$,
something which has not yet been done. It would also show that the full Euler equations possess solutions
which behave like elastically scattering solitary waves.

The neutral, orbital stability of the multi-soliton solutions 
of the KdV equation has been shown by Maddocks and Sachs \cite{madsachs}
based on the observation that the $n$-soliton can be obtained by minimizing
the $n^{th}$ conservation law of the KdV equation subject to the constraints that
the first $n-1$ integrals of the motion are held fixed. On the other hand,
the asymptotic stability of the solitary wave has been demonstrated by Pego and Weinstein
\cite{PW2}, based on a spectral analysis of the linearized Korteweg-deVries
equation. General arguments, based on the
integration of the Korteweg deVries equation by the inverse scattering method, imply that
the n-soliton solution is in some sense asymptotically stable. If so, such a result would
be based on the analysis of the linearized KdV equation at the n-soliton solution.

In this paper we analyze the linearized KdV equation at the multi-soliton
solution. Such an
investigation
was begun by Sachs \cite{Sa}, who constructed a representation of the
linearized
KdV equations using the completeness of the squared eigenfunctions
of the associated Schr\"odinger operator. We extend his analysis here
in the case of the $n$-soliton solution to
obtain estimates in norms suitable for studying perturbations of the
KdV equation.

In the case of the solitary wave, one may work in a reference frame moving with the wave; the result is that the linearized operator has time independent coefficients, and the methods of classical spectral theory of linear operators can be applied. In the
 multi-soliton case this can no longer be carried out, and we are forced to consider time dependent operators. Nevertheless, it is possible to formulate an analog of the single-soliton case, as follows. 

The $n$-soliton solution is a $2n$ parameter family of solutions of the KdV equation. Differentiation with respect to those parameters yields a $2n$-dimensional subspace of solutions of the homogeneous linearized KdV equation which
decay exponentially in $x$ for fixed time. This subspace plays the role of the kernel of the infinitesimal generator in the case of an evolution equation with time-independent coefficients, as occurs when the KdV equation is linearized about the solitary 
wave. The presence of this ``kernel'' leads to $2n$ solvability conditions.

We introduce the space
$\WW$ consisting of functions of the form
$$
u=\sum_{j=1}^n f_j(x-4\o_j^2t)+R(x,t);
$$
 the $f_j$ are ``solitary''-like wave forms which decay exponentially as $x\to \pm\infty$; $f_j$  and $R(x,t)$ are
analytic in a strip in the complex $x$ plane; and $R$  decays exponentially
in time as $t\to \infty$, uniformly in $x$. The class $\WW$ is closed under differentiation and multiplication, an important property when working with nonlinear equations. 
The main purpose of this paper is to invert the KdV equation
linearized at the $n$-soliton solution \reff{n-sol} in $\WW$.

In  \S\ref{sec2} we prove that the $n$-soliton solution belongs to $\WW$. In fact, for the $n$-solitons, $R$ decays exponentially as $|x|\to \infty$ as well.   
This exponential decay is not preserved under
inversion of the linearized KdV equation.
Nevertheless,  by considering weighted spaces, we can show that 
the exponential decay as $x\to \infty$ is  preserved.
In \S\ref{sec3} we derive
the properties of the wave functions of the associated Schr\"odinger operator. 
In \S\ref{sec4} we give a proof of Sachs' completeness theorem, including an
extension to a completeness theorem for the squared eigenfunctions themselves. 

In \S\ref{sec5} we construct the propagator of the linearized KdV equation. 
The basic estimates on the propagator are obtained in \S\S \ref{sec6},\ref{sec8}. In \S\ref{sec6} we obtain estimates in a
Hilbert space of functions analytic in a strip containing the real $x$-axis. 
In \S\ref{sec8} we obtain estimates in a weighted norm, analogous to the estimates 
in \cite{PW1}. In \S\ref{sec7} we discuss the inversion of the 
linearized KdV equation in the space $\WW$ with suitable linear solvability constraints.

The perturbation scheme of the Euler equations which
leads formally to the KdV approximation loses derivatives,
whereas
the inversion of the linearized KdV equation is neutral: it is bounded
in
$L_2$, but gains no derivatives. 
In a situation such as \cite{PW1} the loss of one derivative was
compensated by the use of weighted norms.
Pego and Weinstein used
a global existence theorem for the generalized KdV equation proved by Kato \cite{Ka}, together
with estimates in weighted norms to gain minimal regularity and
decay in time. The weighted norms exact a toll, however; they do not
give rise to Banach algebras, and so are difficult to work with in
nonlinear
problems. 

The loss of derivatives is more severe in
the case of the Euler equations, and one cannot expect the method
in \cite{PW1} to work. 
It seems probable that some form of hard-implicit
function theorem will be needed, 
such as that described by Moser \cite{Mo} or Scheurle \cite{Sc}.
A natural space to work in is the space of functions analytic in a
strip
in the complex $x$ plane. 
\\[2mm] 
{\bf Acknowledgement}
This research was carried out while the second author 
was a Humboldt Preistr\"ager in Stuttgart Universit\"at, 1995-96. He 
greatfully acknowledges the generous support of the Humboldt Stiftung for the
opportunity to study in Germany.
Both authors would like to thank their friend and colleague, 
Klaus Kirchg\"assner, for his constant support and encouragement in this 
work.

\section{Asymptotics of the $n-$soliton solution}
\label{sec2}

We study in this section the asymptotic properties of the $n$-soliton solution
\reff{n-sol}
of the KdV equation, as $t\to \infty$.
The $n$-soliton solution is a function of $2n$ parameters $\o_1,\dots,\o_n$,
$\a_1,\dots,\a_n$. Throughout this paper we fix 
$0<\o_1<\dots <\o_n$ and $\a_1,\dots,\a_n$.
The speeds of the individual solitary waves are $4\o_1^2<\dots <4\o_n^2$;
the $\a_1,\dots,\a_n$ are called the phases. The determinant in \reff{n-sol}
is called the  tau function of order $n$ and is denoted by $\tau$.

\btheo{nasym} The $n$-soliton solution of the KdV equation can be written in
the form
\beq{ms}
u(x,t)=-2\sum_{j=1}^n \o_j^2 \sech^2(\theta_j+\gamma_j)
-2\frac{d^2}{dx^2}
\log(1+R),
\end{equation}
where 
$$
\gamma_n=\fr12\log(2\o_n),\quad \gamma_j=\fr12\log(2\o_j)+
\sum_{k=j+1}^n \log\left(\fr{\o_k+\o_j}{\o_k-\o_j}\right),\quad
1\leq j\leq n-1,
$$
 and
\beq{Rest}
\sup_{x,t>0} |\cosh(a x)R(x,t)| \leq Ce^{-b t},
\end{equation}
for some  $a,b>0$ and some positive constant $C$.  
\etheo

A similar result is true as $t\rightarrow -\infty$, 
but with different phase shifts $\gamma_j$.
\bigskip

\Proof
We will prove that the tau function of order $n$ can be factored
$$
\tau(\theta_1,\dots,\theta_n)=2e^{-(\theta_n+\g_n)}\cosh(\theta_n +\g_n)
\tau(\theta_1+\b^n_1,\dots,\theta_{n-1}+\b^n_{n-1})(1+R_n),
$$
where $\tau(\theta_1+\b^n_1,\dots,\theta_{n-1}+\b^n_{n-1})$ is the tau function 
of order $n-1$,
$$
\b^n_j=\log\left(\fr{\o_n+\o_j}{\o_n-\o_j}\right)>0,
$$
and
 $R_n$ satisfies \reff{Rest}. Then by induction,
$$
\tau=\prod_{j=1}^n2e^{-(\theta_j+\g_j)}\cosh(\theta_j+\g_j)(1+R_j),
$$
where each of the $R_j$ satisfies \reff{Rest}. The result then follows for the KdV solution
$u$ upon taking the second logarithmic derivative and letting
\beq{Rprod}
1+R=\prod_{j=1}^n (1+R_j).
\end{equation}

We begin by writing 
$$
\tau=\det(I+C_n),
\qquad
C_{n}=\left[\frac{e^{-(\theta_j+\theta_k)}}{\o_j+\o_k}\right]_{1\leq j,k\leq n}.
$$
 As observed in
\cite{GGKM} $\tau$  can be expanded as a sum of all the
principal minors of $C_n$. Moreover, each of these principal minors is of the same form. We may write
$$
C_n=\Lambda_n K_n \Lambda_n,
\qquad
\Lambda_n=\text{diag} (e^{-\theta_1},\dots , e^{-\theta_n}),
\qquad
K_{n}=\left[\frac{1}{\o_j+\o_k}\right]_{1\leq j,k\leq n}.
$$
Thus,
\begin{multline}\label{tau}
\tau (\theta_1,\dots,\theta_n) =\\[4mm]
 1+\sum_{j=1}^n \frac{e^{-2\theta_j}}{2\o_j}+\sum_{1\leq j<k\leq n}e^{-2(\theta_j+\theta_k)}K^{(2)}_{jk}
+\dots +e^{-2(\theta_1+\dots +\theta_n)}K^{(n)},
\end{multline}
where $K^{(\ell)}_{j_1\dots j_\ell}$ is the $\ell\times\ell$ principal minor on the rows 
$j_1,\dots, j_\ell$.

\blem{reduct} We have
$$
\tau(\theta_1,\dots ,\theta_n)=\tau(\theta_1,\dots,\theta_{n-1})+
e^{-2(\theta_n+\gamma_n)}
\tau(\theta_1+\b^n_1,\dots,\theta_{n-1}+\b^n_{n-1}).
$$
\elem

\Proof
In \reff{tau} the only terms which do not contain the factor 
$e^{-2\theta_n}$ are the principal minors of the matrix $C_{n-1}$, so we have the 
factorization
\begin{multline*}
\tau(\theta_1,\dots,\theta_n)=\tau(\theta_1,\dots,\theta_{n-1})+
e^{-2\theta_n}\Big(\fr{1}{2\o_n}+\sum_{j=1}^{n-1}e^{-2\theta_j}K_{jn}^{(2)}
\\[4mm]
 +\sum_{1\leq j<k\leq n-1}e^{-2(\theta_j+\theta_k)}K_{jkn}^{(3)}+
\dots+e^{-2(\theta_1+\dots +\theta_{n-1})}K^{(n)}\Big).
\end{multline*}

The following formula has been proved in \cite{GGKM}, p.121: 
\beq{Kn}
\det K_n=\fr{1}{2\o_n}\prod_{j=1}^{n-1}
\left(\fr{\o_n-\o_j}{\o_n+\o_j}\right)^2 \det K_{n-1},
\end{equation}
so
$$
K_{j_1\dots j_ln}^{(l+1)}=\fr{1}{2\o_n} \prod_{\a=1}^{l}
\left(\fr{\o_n-\o_{j_\a}}{\o_n+\o_{j_\a}}\right)^2 K_{j_1\dots j_l}^{(l)}.
$$

Then 
\begin{align*}
\tau(\theta_1,\dots,\theta_n)=&
\tau(\theta_1,\dots,\theta_{n-1})+\fr{e^{-2\theta_n}}{2\o_n}
\left(1+\sum_{j=1}^{n-1}\left(\fr{\o_n-\o_j}{\o_n+\o_j}\right)^2 
e^{-2\theta_j}K_{j }^{(1)}\right.\\[4mm]
&\hskip 10mm + \sum_{1\leq j<k\leq n-1}
\left(\fr{\o_n-\o_j}{\o_n+\o_j}\right)^2 
\left(\fr{\o_n-\o_k}{\o_n+\o_k}\right)^2
 e^{-2(\theta_j+\theta_k)}K_{jk}^{(2)}\\[4mm]
&\hskip 10mm \left. +\dots+\prod_{k=1}^{n-1}\left(\fr{\o_n-\o_k}{\o_n+\o_k}\right)^2
e^{-2\theta_k}K^{(n-1)}\right)\\[4mm]
&=\tau(\theta_1,\dots,\theta_{n-1})+
e^{-2(\theta_n+\gamma_n)}
\tau(\theta_1+\b^n_1,\dots,\theta_{n-1}+\b^n_{n-1}),
\end{align*}
since
$$
e^{-2\gamma_n}=\fr{1}{2\o_n},\quad 
e^{-2\b^n_j}=\left(\fr{\o_n-\o_j}{\o_n+\o_j}\right)^2 , 
\quad 1\leq j\leq n-1.
$$
\qed

From this lemma, we can write the $\tau$-function in the form,
$\widetilde{\theta_n}=\theta_n+\gamma_n$,
\begin{align*}
\tau(\theta_1,\dots,\theta_n)
=& 2e^{-\widetilde{\theta_n}}\cosh\widetilde{\theta_n} \Big( \fr{e^{\widetilde{\theta_n}}}{2\cosh\widetilde{\theta_n}}\tau(\theta_1,\dots,\theta_{n-1})\\[4mm]
&\hskip 10mm +\fr{e^{-\widetilde{\theta_n}}}
{2\cosh\widetilde{\theta_n}}\tau(\theta_1+\b^n_1,\dots,
\theta_{n-1}+\b^n_{n-1})\Big)\\[4mm]
=& 2e^{-\widetilde{\theta_n}}\cosh\widetilde{\theta_n} \Big( \fr{1+\tanh\,\widetilde{\theta_n}}{2}
\tau(\theta_1,\dots,\theta_{n-1})\\[4mm]
& \hskip 10mm +\fr{1-\tanh\,\widetilde{\theta_n}}{2}
\tau(\theta_1+\b^n_1,\dots,\theta_{n-1}+\b^n_{n-1})\Big)\\[4mm]
=& 2e^{-\widetilde{\theta_n}}\cosh\widetilde{\theta_n}\, \tau(\theta_1+\b^n_1,\dots,\theta_{n-1}+\b^n_{n-1})(1+R_n),
\end{align*}
where
\begin{align*}
1+R_n=&\fr{1-\tanh\,\widetilde{\theta_n}}{2}+\fr{1+\tanh\,\widetilde{\theta_n}}{2}
\fr{\tau(\theta_1,\dots,\theta_{n-1})}{\tau(\theta_1+\b^n_1,\dots,\theta_{n-1}+\b^n_{n-1})}\\[6mm]
=& 1+\fr{1+\tanh\,\widetilde{\theta_n}}{2}\left[
\fr{\tau(\theta_1,\dots,\theta_{n-1})}{\tau(\theta_1+\b^n_1,\dots,\theta_{n-1}+\b^n_{n-1})} -1
\right].
\end{align*}
 
To complete the proof of the theorem we have to show that $R_n$ satisfies 
\reff{Rest}. Remark first that
$$
1\leq \tau(\theta_1+\b^n_1,\dots,\theta_{n-1}+\b^n_{n-1})\leq \tau(\theta_1,\dots,\theta_{n-1}),
$$
and that the ratio 
$$
\fr{\tau(\theta_1,\dots,\theta_{n-1})}{\tau(\theta_1+\b^n_1,\dots,\theta_{n-1}+\b^n_{n-1})}
$$
is bounded for all values of $\theta_1,\dots,\theta_{n-1}$ (it is the ratio of two 
positive polynomials of $e^{-\theta_1},\dots,e^{-\theta_{n-1}}$ having the same degree).

Then
\begin{align*}
0\le & R_n
=\fr{1+\tanh\,\widetilde{\theta_n}}{2}
\left[ \fr{\tau(\theta_1,\dots,\theta_{n-1})}{\tau(\theta_1+\b^n_1,\dots,\theta_{n-1}+\b^n_{n-1})}
-1\right] \\[4mm]
\le & C_1\fr{1+\tanh\,\widetilde{\theta_n}}{2}
\leq C_1e^{2\widetilde{\theta_n}},
\end{align*}
and the required inequality is immediate if $x\leq 0$.

For $0\leq x\leq 2(\o_{n-1}^2
+\o_n^2)t$, we find, for all $a\in\R$,
\begin{align*}
0\leq R_n
\le & C_1e^{2\widetilde{\theta_n}}=
 \fr{C_1}{2\o_n}e^{-2\o_n\a_n}
e^{-a x+(2\o_n+a)x-8\o_n^3t}\\[4mm]
\le& \fr{C_1}{2\o_n}e^{-2\o_n\a_n}
e^{-a x+[2(\o_{n-1}^2+\o_n^2)(2\o_n+a)-8\o_n^3]t}\\[4mm]
\le& \fr{C_1}{2\o_n}e^{-2\o_n\a_n}
e^{-a x-[4\o_n(\o_n^2-\o_{n-1}^2)-2a(\o_n^2+\o_{n-1}^2)]t},
\end{align*}
Choose $a>0$, 
$
a<2\o_n(\o_n^2-\o_{n-1}^2)/(\o_n^2+\o_{n-1}^2),
$
 and \reff{Rest} follows with 
$$
b=4\o_n(\o_n^2-\o_{n-1}^2)-2a(\o_n^2+\o_{n-1}^2).
$$ 

Now, for $x\geq 2(\o_{n-1}^2+\o_n^2)t$
we have $\theta_j=\o_j(x-\a_j-4\o_j^2t)>-\o_j\a_j$, if $j=1,\dots, n-1$; and,
for $0<a<2\o_1$,
\begin{align*}
0\le & R_n
\le \tau(\theta_1,\dots,\theta_{n-1})-\tau(\theta_1+\b^n_1,\dots,\theta_{n-1}+\b^n_{n-1})
\\[2mm]
=&\sum_{j=1}^{n-1} (1-e^{-2\b^n_j}) \frac{e^{-2\theta_j}}{2\o_j}
+\sum_{1\leq j<k\leq n-1}(1-e^{-2(\b^n_j+\b^n_k)})e^{-2(\theta_j+\theta_k)}K^{(2)}_{jk}\\[4mm]
&\hskip 10mm +\dots
 +(1-e^{-2(\b^n_1+\dots+\b^n_{n-1})})
e^{-2(\theta_1+\dots +\theta_{n-1})}K^{(n-1)}\\[4mm]
\le&C_2\sum_{j=1}^{n-1}e^{-2\theta_j}
\leq C_3\sum_{j=1}^{n-1}
e^{-2\o_jx+8\o_j^3t}=C_3\sum_{j=1}^{n-1}
e^{-a x -(2\o_j-a)x+8\o_j^3t}
\end{align*}
\begin{align*}
\le& C_3\sum_{j=1}^{n-1}
e^{-a x -2(\o_{n-1}^2
+\o_n^2)(2\o_j-a)t+8\o_j^3t}
\\[2mm]
\le&
 C_3\sum_{j=1}^{n-1}
e^{-a x-[4\o_j(\o_n^2+\o_{n-1}^2-2\o_1^2)-2a(\o_1^2+\o_n^2)]t}.
\end{align*}

 As before \reff{Rest} follows if  
$0<a<2\o_j(\o_n^2+ \o_{n-1}^2-2\o_j^2)/(\o_n^2+ \o_{n-1}^2) $.\qed
This theorem shows that the $n$-soliton
solution is asymptotic to a sum of $n$
travelling solitary waves plus a remainder term that decays exponentially
fast to zero as $t\rightarrow \infty$, uniformly in $x$.

In the case of 2-solitons, a
simple
computation shows that
$$
\tau=1+\fr{e^{-2\theta_1}}{2\o_1}
+\fr{e^{-2\theta_2}}{2\o_2}+\fr{1}{4\o_1\o_2}
\left(\fr{\o_2-\o_1}{\o_2+\o_1}\right)^2e^{-2(\theta_1+\theta_2)}.
$$
 We may factor  $\tau$  as
$$
\tau=2e^{-(\theta_2+\gamma_2)}\cosh (\theta_2+\gamma_2)\tau_1,
$$
where $\gamma_2=\frac12\log (2\o_2)$ and
$$
\tau_1=1+\frac{1}{4\o_1} e^{-2\theta_1}(1+\tanh(\theta_2+\gamma_2))
+\frac{1}{4\o_1}\left(\fr{\o_2-\o_1}{\o_2+\o_1}\right)^2
e^{-2\theta_1}(1-\tanh(\theta_2+\gamma_2)).
$$

As $t\rightarrow \infty$,
\beq{lims}
(1-\tanh(\theta_2+\gamma_2))\rightarrow 2,
\qquad
(1+\tanh(\theta_2+\gamma_2))\rightarrow 0;
\end{equation}
hence
$$
\tau_1\sim 1+\frac{1}{2\o_1} \left(\fr{\o_2-\o_1}{\o_2+\o_1}\right)^2
e^{-2\theta_1}
=2e^{-(\theta_1+\gamma_1)}\cosh( \theta_1+\gamma_1),
$$
where 
$$
\gamma_1=\fr{1}{2}\log (2\o_1)+\log \left(\fr{\o_2+\o_1}{\o_2-\o_1}\right).
$$
This leads, ultimately, to the factorization
$$
\tau=4e^{-(\theta_1+\theta_2+\gamma_1+\gamma_2)}
\cosh(\theta_2+\gamma_2) \cosh(\theta_1+\gamma_1) (1+R),
$$
with 
\beq{R}
R=\fr{\o_1\o_2}{(\o_2-\o_1)^2}(1+{\rm tanh}(\theta_2+\gamma_2))
(1-{\rm tanh}(\theta_1+\gamma_1)).
\end{equation}

Hence,
the 2-soliton solution of the KdV equation can be written
$$
u(x,t)=-2\o_1^2{\rm sech}^2(\theta_1+\gamma_1)
-2\o_2^2{\rm sech}^2(\theta_2+\gamma_2)-2\fr{d^2}{dx^2}\log (1+R).
$$

The diagram below shows the {\it negative} of the
two-soliton solution \reff{n-sol}, with $\o_1,\,\o_2=.5,\,.75$, and
$\a_1=\a_2=0$ (solid lines). The negative is the leading term
in the approximation of the free surface in
the full Euler equations.
The first diagram shows the two-soliton solution in the middle of the
 interaction.
The diagram on the right shows the two-soliton solution after the interaction.
The boxed line shows the superposition of two solitary ($\sech^2$) waves.
These two solitary waves fit the two-soliton solution exactly before
the interaction. The displacement is the scattering
of the solitary waves due to the non-linear interaction.
\vs
\centerline{\begin{picture}(0,0)%
\epsfig{file=2sol.pstex}%
\end{picture}%
\setlength{\unitlength}{0.00066700in}%
\begingroup\makeatletter\ifx\SetFigFont\undefined
\def\x#1#2#3#4#5#6#7\relax{\def\x{#1#2#3#4#5#6}}%
\expandafter\x\fmtname xxxxxx\relax \def\y{splain}%
\ifx\x\y   
\gdef\SetFigFont#1#2#3{%
  \ifnum #1<17\tiny\else \ifnum #1<20\small\else
  \ifnum #1<24\normalsize\else \ifnum #1<29\large\else
  \ifnum #1<34\Large\else \ifnum #1<41\LARGE\else
     \huge\fi\fi\fi\fi\fi\fi
  \csname #3\endcsname}%
\else
\gdef\SetFigFont#1#2#3{\begingroup
  \count@#1\relax \ifnum 25<\count@\count@25\fi
  \def\x{\endgroup\@setsize\SetFigFont{#2pt}}%
  \expandafter\x
    \csname \romannumeral\the\count@ pt\expandafter\endcsname
    \csname @\romannumeral\the\count@ pt\endcsname
  \csname #3\endcsname}%
\fi
\fi\endgroup
\begin{picture}(8124,3051)(889,-3400)
\put(2701,-3361){\makebox(0,0)[b]{\smash{\SetFigFont{12}{14.4}{rm}$t=0$}}}
\put(7201,-3361){\makebox(0,0)[b]{\smash{\SetFigFont{12}{14.4}{rm}$t=10$}}}
\end{picture}

}
\vs
\btheo{multi} 
The $n$-soliton solution \reff{n-sol} is analytic in
the strip
$|\Im x|<\pi/ 2\o_n$. 
Moreover, \reff{ms} is valid for $|\Im x|<\pi/2\o_n$ and \reff{Rest} holds for 
$|\Im x|\leq\eta_0$, for any $0<\eta_0<\pi/2\Omega$, where 
$\Omega=\max\{2\sum_{j=1}^{n-1}\o_j,\o_n\}$. 
The constant $C$ in \reff{Rest} depends only on $\eta_0$.
\etheo

\Proof The Gel'fand-Levitan-Marcenko (GLM) equation for the $n$-soliton solutions is (cf. \cite{GGKM})
\beq{glm}
K(x,y,t)+f(x+y,t)+\inxp K(x,s,t)f(s+y,t)\,ds=0,
\end{equation}
where
$$
f(\xi,t)=\sum_{j=1}^n e^{-\o_j\xi+8\o_j^3t+2\o_j\a_j}.
$$ 
We need to prove its invertibility for  complex $x$.
The integral in \reff{glm} should be understood as a complex
integral over $\Gamma=\{ z\in\C  :  \Re z>\Re x, \Im z=\Im x\}$.
The $n$-soliton solution is then given by
$$
u(x,t)=-2\fr{d}{dx}K(x,x,t).
$$

Since \reff{glm} is a Fredholm integral equation, 
its solvability follows from uniqueness. We first
make a transformation and rewrite the homogeneous equation as
\beq{homglm}
K_{x,t}(y)+\int_0^\infty K_{x,t}(s)f(s+y+2x,t)\,ds=0
\qquad
K_{x,t}(y)=K(x,x+y,t).
\end{equation}
Now it is clear that $x$ appears analytically in the equation, 
and so the solutions of \reff{glm}
will be analytic in $x$ wherever we can prove uniqueness.

To prove uniqueness, multiply \reff{homglm} by $\overline{K_{x,t}(y)}$ 
and integrate
 over
$(0,\infty)$. This leads to
\begin{align*}
0=& \int_0^\infty |K_{x,t}(y)|^2\,dy+\int_0^\infty\int_0^\infty 
f(s+y+2x,t)K_{x,t}(s)\overline{K_{x,t}(y)}\,dy ds\\
=&\int_0^\infty |K_{x,t}(y)|^2\,dy+\sum_{j=1}^n 
e^{-2\o_j x+8\o_j^3t+2\o_j\a_j}
\big| \int_0^\infty e^{-\o_js}K_{x,t}(s)\,ds\big|^2.
\end{align*}

The above expression can vanish only when the convex hull of the
complex numbers $e^{-2i\o_j \eta}$, where $x=\xi+i\eta$, contains the
negative real axis, hence only when $|\eta|\geq \pi/2\o_n$.
The first part of the theorem is now proved.

The waveform $\sech^2 (\o_j x)$ is analytic in the
strip $|\Im x|< \pi/2\o_j$: it has poles on the imaginary axis at the
points $x= i(\frac\pi2+k\pi)/\o_j$, $k\in\R$. Hence the left hand side and 
the sum in the right hand side of \reff{ms} are analytic in the strip
$|\Im x|< \pi/2\o_n$, so \reff{ms} holds in this strip.

To obtain the estimate \reff{Rest}  remark that  the arguments 
in the proof of Theorem \ref{nasym} 
remain valid as long as all the exponential terms
$e^{-2(\theta_1+\dots +\theta_j)}$, $j=1,\dots, n-1$ have positive real
parts,
and $1+\tanh\,\theta_n$ is uniformly bounded. These properties hold in
any strip $|\Im x|\leq\eta_0$, for $0<\eta_0<\pi/2\Omega$.
\qed
\\
{\bf Remark.}  The argument  in the last part of the proof of this theorem 
implies also the analyticity of the $n$-soliton solution in the strip $|\Im x|<
\pi/2\Omega$. However, this  result is weaker  than the one above.  
We conjecture that \reff{Rest} holds in fact in any strip           
$|\Im x|\leq \eta_0$, for $\eta_0<\pi/2\o_n$.

\section{Wave functions}
\label{sec3}

We study in this section some of the properties of the wave functions
of the Schr\"odinger operator
\beq{1}
(D^2+k^2-u(x,t))\psi(x,k)=0,
\end{equation}
when $u$ is an $n$-soliton solution of the KdV equation
\beq{kdv}
u_t+ u_{xxx}-6 uu_x=0.
\end{equation}

The Lax pair for this equation is
\beq{lpair}
L=- D^2+ u,
\qquad
B=-4D^3+3( uD+Du).
\end{equation}
Substituting $u=\vphi_x$ into the above form of the KdV equation, we obtain
$$
\vphi_{x,t}+ \vphi_{xxxx}-6
 \vphi_x\vphi_{xx}=0,
$$
or after integration, the {\it potential KdV} equation
\beq{pkdv}
\vphi_t+\vphi_{xxx}-3\vphi_x^2=0.
\end{equation}

The linearized KdV  equation is
\beq{lkdv}
v_t+v_{xxx}-6(uv)_x=0,
\end{equation}
while the linearization of \reff{pkdv} is
\beq{lpkdv}
\psi_t+ \psi_{xxx}-6u\psi_x=0.
\end{equation}
Thus the linearized potential KdV and the KdV equations are formal adjoints
of one another.

\blem{l6}  Let $q$ be a solution of \reff{kdv}, and let $f_1,\ f_2$ be any pair of
solutions of $Lf=k^2f,\ \ f_t=Bf,$
where $L$ and $B$ are given in \reff{lpair}. Then $\psi(x,t,k)=(f_1f_2)(x,t,k)$ satisfies the linearized potential KdV equation
\reff{lpkdv}, and $\psi_x$ satisfies the linearized KdV equation \reff{lkdv}.
\elem

\Proof 
Straightforward calculation, \cite{Sa}.\qed

The asymptotic behaviour 
of the wave functions is
determined by the large $x$ behaviour of the Lax pair $L$ and $B$. It is
thus given by 
the simultaneous equations 
$$
-D^2\psi=k^2\psi, \qquad \psi_t= -4D^3\psi
$$
This leads to 
\beq{phi}
\vphi_+(x,t,k)=m(x,t,k)e^{-i(kx+4k^3t)},
\qquad
\psi_+(x,t,k)=r(x,t,k)e^{i(kx+4k^3t)},
\end{equation}
The reduced wave functions $m(x,t,k),\ r(x,t,k)$ are
normalized by 
\beq{norm}
\lim_{x\to -\infty}m(x,t,k)=1,\quad \lim_{x\to\infty}
r(x,t,k)=1.
\end{equation}

\blem{mr} The reduced wave functions $m(x,t,k)$ and $r(x,t,k)$ are analytic in
$|\Im x|< 2\pi/\Omega$, and $\Im k>0$.
Moreover, $m$, $r$ and their derivatives with respect to $x$ are uniformly
bounded for $|\Im x|\leq \eta_0$, $t\in\R$,  $\Im k\geq\e$,
for any $0<\eta_0< 2\pi/\Omega$ and $\e>0$.
\elem

\Proof
The reduced wave functions $m(x,t,k),\ r(x,t,k)$ are
obtained as solutions of Volterra integral equations
\begin{align}
\quad m(x,t,k)=&1-\inmx \frac{1-e^{2ik(x-y)}}{2ik}u(y,t)m(y,t,k)\,dy ,
 \nonumber\\[6mm]
\quad r(x,t,k)=&1-\inxp \frac{1-e^{-2ik(x-y)}}{2ik}u(y,t)r(y,t,k)\,dy .
\label{volterra}
\end{align}
From Theorem \ref{multi} it follows that the integrals 
$$
\inmx \frac{|1-e^{2ik(x-y)}|}{2|k|}|u(y,t)|\,dy ,\quad
\inxp \frac{|1-e^{-2ik(x-y)}|}{2|k|}|u(y,t)|\,dy,
$$
 are  uniformly bounded
for $|\Im x|\leq \eta_0$, $t\in\R$, and $\Im k\geq\e$,
for any $0<\eta_0< 2\pi/\Omega$ and $\e>0$. Then, 
the existence, boundedness, and analyticity of $m,\ r$ follow easily
from standard arguments for
Volterra integral equations. 

The uniform boundedness of the derivatives of $m$ and $r$
with respect to $x$ follows from the uniform boundedness
of $m$ and $r$ by using the 
Volterra integral equations \reff{volterra} in 
differentiated form, e.g.
$$
m_x(x,t,k)=
\inmx e^{2ik(x-y)}u(y,t)m(y,t,k)\,dy.
$$
\qed

The conjugate wave functions are 
$$
\vphi_-(x,t,k)=\overline{\vphi_+(x,t,\bar k)},
\qquad
\text{and} \psi_-(x,t,k)=\overline{\psi_+(x,t,\bar k)}.
$$
Since the $n$-soliton solutions are reflectionless potentials 
they
satisfy
\beq{scat}
\vphi_+=a(k)\psi_-,\qquad \psi_+=a(k)\vphi_-,
\end{equation}
where
$$
a(k)=\frac {W(\vphi_+(x,t,k),\, \psi_+(x,t,k))}{2ik}.
$$

For the $n$-soliton solution of the KdV equation both wave functions
$\vphi_+$ and $\psi_+$ are in fact meromorphic functions of $k$ with simple poles at 
$-i\o_1,\dots,-i\o_n$. 
{From} \reff{phi}
and \reff{norm} we deduce
the following principal parts expansion of the wave functions:
\begin{align}
\vphi_+(x,t,k)={e^{-i(kx+4k^3t)}
\left[1+\sum_{j=1}^n\frac{n_j(x,t)}{\o_j-ik}\right]},
\\[6mm]
\psi_+(x,t,k)={e^{i(kx+4k^3t)}
\left[1+\sum_{j=1}^n\frac{p_j(x,t)}{\o_j-ik}\right],}\label{wf}
\end{align}
where $n_j,\,p_j$ are the residues of the reduced wave functions at the poles $k=-i\o_j$.
Then, from \reff{wf} and from Lemma \ref{mr} follows.

\btheo{wavef}
The reduced wave functions $m$, $r$ 
and their derivatives with respect to $x$ are 
uniformly bounded in
$|\Im x|\leq \eta_0$, $t\in\R$,  
$\Im k\geq 0$, for any $0<\eta_0<2\pi/\Omega$.
\etheo

We define the eigenfunctions corresponding to the bound states $k=i\o_j$ by
\beq{phim}
\vphi_j(x,t)=\vphi_+(x,t,i\o_j),\quad
\psi_j(x,t)=\psi_+(x,t,i\o_j).
\end{equation}
They can be calculated explicitly by solving the GLM equation, cf.
\cite{GGKM}.
In \cite{GGKM} the bound state eigenfunctions are normalized differently, namely so that $\int \psi_j^2=1$; this leads to their constraint (2.13). Comparing (2.13) in \cite{GGKM} and \reff{norm}, we see that 
$$
\psi_j^{[GGKM]} = d_j\psi_j,
\qquad
d_j=e^{\o_j\a_j}.
$$

Then, according to (3.3) in \cite{GGKM}, 
the kernel $K$ in the GLM equation is of the form
\beq{eK}
K(x,y,t) =-\sum_{j=1}^n \cj d_j\psi_j(x,t)e^{-\o_jy},
\end{equation}
with $d_j $ the constants  determined above.
 The substitution of \reff{eK} into \reff{glm} leads
to a linear system for $\psi_j$:
\beq{spsi}
d_j\psi_j+\sum_{m=1}^n\fr{e^{-(\theta_j+\theta_m)}}{\o_j+\o_m}
d_m\psi_m=e^{-\theta_j}.
\end{equation}
Furthermore, from Theorem 3.4 in \cite{GGKM}
we deduce the following relation between the $n$-soliton solution and
the bound state eigenfunctions
\beq{ruw}
u(x,t)=-4\sum_{j=1}^n\o_je^{2\o_j\a_j}\psi_j^2(x,t).
\end{equation}

We now look for the wave function $\psi_+$. For this we determine
$p_j$ in \reff{wf}  in terms of $\psi_j$. 
{From} \reff{wf} and \reff{phim} we find
$$
\psi_j(x,t)=e^{-\o_jx+4\o_j^3t}\left[1+
\sum_{m=1}^n\fr{p_m(x,t)}{\o_m+\o_j}\right].
$$
Multiply this equation by $d_j=e^{\o_j\a_j}$, and recall that $\theta_j=\o_j(x-\a_j-4\o_j^2t)$. Comparing the resulting system with \reff{spsi} we see that
\beq{pj}
p_j(x,t)=-e^{-\theta_j}d_j\psi_j(x,t)=
-e^{-\o_jx+4\o_j^3t+2\o_j\a_j}\psi_j(x,t).
\end{equation}
Hence
\beq{fpsi}
\psi_+(x,t,k)=e^{i(kx+4k^3t)}
\left[1-\sum_{j=1}^n\frac{e^{-\o_jx+4\o_j^3t
+2\o_j\a_j}}{\o_j-ik}\psi_j(x,t)\right].
\end{equation}

We can also obtain an explicit formula for $\vphi_+$ from \reff{scat}. 
For this we need $a(k)$. 
First note that 
$$
\vphi_-(x,t,k)=\overline{\vphi_+(x,t,\bar{k})}\sim e^{i(kx+4k^3t)},
\qquad \text{as}\quad
x\rightarrow -\infty.
$$
Equations \reff{scat} and \reff{fpsi} then imply  
\beq{ak1}
a(k)=\lim_{x\to -\infty}e^{-i(kx+4k^3t)}\psi_+(x,t,k)=1-\lim_{x\to-\infty }\sum_{j=1}^n\frac{e^{\o_j\a_j-\theta_j}}{\o_j-ik}\psi_j(x,t).
\end{equation}
The limiting values of $e^{-\theta_j}\psi_j$ as $x\to -\infty$ can be 
determined from 
\reff{spsi}. Multiply  \reff{spsi} by $e^{\theta_j}$ 
and note that $e^{\theta_j}$ tends to zero exponentially 
as $x\to-\infty$. Moreover, $\psi_m$ is the eigenfunction 
corresponding to the eigenvalue $-\o_m^2$ and 
is in $L^2(\mathbb R)$. (In fact, it also tends 
to zero exponentially as
$x\to \pm\infty$.) Hence, these limits satisfy a linear system with
constant coefficients, so
$$
a(k)=1-\sum_{j=1}^n\fr{a_j}{\o_j-ik}=\fr{P_n(k)}{{\prod_{j=1}^n}(\o_j-ik)},
$$
with $P_n(k)$ a polynomial of $k$ of degree $n$. But $a(i\o_j)=0$, for $j=1,\dots,n$,
so 
$$
P_n(k)=\a\prod_{j=1}^n(\o_j+ik).
$$
The constant $\a$ is determined from the asymptotic properties of 
$a(k)$ for large $k$. From \reff{ak1} we find
$$
\lim_{k\to\infty}a(k)=1,
$$
so $\a=(-1)^n$, and we conclude
\beq{ak}
a(k)=\prod_{j=1}^n\fr{k-i\o_j}{k+i\o_j}.
\end{equation}

Now, \reff{scat} yields
\beq{fphi}
\vphi_+(x,t,k)=
e^{-i(kx+4k^3t)}\prod_{j=1}^n\fr{k-i\o_j}{k+i\o_j}
\left[1-\sum_{j=1}^n\frac{e^{\o_j\a_j-\theta_j}}{\o_j+ik}
\psi_j(x,t)\right].
\end{equation}

Finally,  we compute the coupling coefficients $c_j$ between 
$\vphi_j$ and $\psi_j$, defined by $\vphi_j=c_j\psi_j.$ 
{From} \reff{fpsi} a straightforward calculation yields
\beq{cm}
c_j=\fr{e^{2\a_j\o_j}}{2\o_j}\prod_{m\neq j}
\fr{\o_j-\o_m}{\o_j+\o_m}.
\end{equation}
Note that, with our choice of normalization of the wave functions, the coupling coefficients 
for the bound state wave functions are time independent!

\section{Completeness theorem}
\label{sec4}

Sachs \cite{Sa} proved a completeness theorem for the derivatives
of
the squared eigenfunctions of the Schr\"odinger operator
\reff{1},
and used this to construct the inverse
operator for the linearized KdV equation. We describe in this section a modification
of Sachs' result that gives a direct extension to a result for the squared eigenfunctions themselves.
For simplicity we assume that the potential $u$ is a multi-soliton solution, though the results are still valid
when $u$  is not a reflectionless potential.

The squared eigenfunctions satisfy a third order linear
homogeneous differential
equation, which we write in the form
\beq{2}
M\vphi=(D^3-2(uD+Du)+4k^2D)\vphi(x,k)=0,
\end{equation}
where $\vphi$ is the product of any pair of solutions of
\reff{1}. The function $u$ here is the $n$-soliton solution
in \S\ref{sec2} and \S\ref{sec3} (below $t$ will appear only as a 
parameter, and for simplicity we do not write the dependence of the functions
on $t$). 
Using the wave functions $\vphi_+(x,k)$ and 
$\psi_+(x,k)$ as a basis of 
solutions of \reff{1} we obtain
$$
\vphi_+^2(x,k), \qquad
\psi_+^2(x,k),
\qquad
\vphi_+(x,k)\psi_+(x,k),
$$
as a basis of solutions of \reff{2}.
 
The completeness theorem is based on the following formal
calculation of the contour integral
of the resolvent
of the operator $M$. Let $\Gamma_R$ be the semi-circle of
radius $R$ in
the upper half plane traversed from -1 to 1.
Formally
\begin{align*}
\lim_{R\to \infty} &\frac{1}{2\pi i}\int_{\Gamma_R}
\,(D^3-2(uD+Du)+4k^2)^{-1}\,8k\,dk\\
=&-\lim_{R\to \infty} \frac{1}{2\pi i}\int_{C_R}
D^{-1}(D^2-2(u+DuD^{-1})-\l)^{-1}\,d\l
\qquad\quad (\l=-4k^2)\\
=&D^{-1},
\end{align*}
where $C_R$ is the circle of radius $4R^2$  in the
$\l$-plane oriented in the counter-clockwise direction.

We construct the inverse of $M$ using as a basis of homogeneous
solutions the above set of squared eigenfunctions.
Since $M$ is skew symmetric, we seek a  kernel 
$\KK(x,y,k)$ satisfying 
$$
M\KK(x,y,k)=\delta(x-y),\qquad
\KK(x,y,k)=-\KK(y,x,k).
$$
Thus
$$
[\KK]=0, \qquad [D_x\KK]=0, \qquad [D_x^2\KK]=1,
$$
where $[\KK]$ denotes the jump of $\KK$ across the singularity $x=y$
(from $x<y$ to $x>y$),
etc. We also require $\KK(x,y,k)$ to be meromorphic in $\Im k>0$ and
bounded as $x\to -\infty$ and $y\to \infty$.

Recall that, for $\Im k>0$,
\begin{align*}
\vphi_+^2(x,k)\to & 0, \quad \text{as}\quad x\to -\infty,\\ 
\psi_+^2(x,k)\to & 0, \quad \text{as}\quad x\to \infty, \\
\vphi_+\psi_+=&a(k)\psi_-\psi_+
\to a(k),\quad \text{as}\quad x\to \infty.
\end{align*}

We try
$$
\KK(x,y,k)=\vphi_+^2(x,k)C_1(y,k)+\vphi_+(x,k)\psi_+(x,k)C_2(y,k),
\qquad
x<y.
$$
The jump conditions across $x=y$ give three equations for
the two unknowns $C_1$ and $C_2$, but one finds that the
equations are consistent. After some computations, we obtain
\beq{3}
\KK(x,y,k)=\vphi_+(x,k)\psi_+(y,k)\fr{R(x,y,k)}{-8k^2a^2(k)},
\qquad
x<y,
\end{equation}
where
\beq{Rxy}
R(x,y,k)=\vphi_+(x,k)\psi_+(y,k)-\vphi_+(y,k)\psi_+(x,k),
\qquad
x<y.
\end{equation}

Based on these considerations, we now prove

\blem{l5} For $f\in L_1(\mathbb R)$ we have
\beq{4}
\int_{x_0}^{x}f(y)\,dy=
\lim_{R\to \infty}\int_{\Gamma_R}\inmp
[\KK(x,y,k)-\KK(x_0,y,k)]f(y)\,dy\,\fr{8k\,dk}{2\pi i}.
\end{equation}
\elem

\Proof 
{From} the asymptotic properties of the wave functions \reff{phi}
it is easy to see that
$$
8k\KK(x,y,k) =\frac{1-e^{2ik(y-x)}}{k}\Big(1+O(\frac{1}{|k|})\Big)
,\quad \text{as} \quad |k|\to \infty,
$$
if $x<y$.
So for $x_1<x_2$,
$$
8k[\KK(x_2,y,k)-\KK(x_1,y,k)] \sim
\begin{cases}
\frac{e^{2ik(y-x_1)}-e^{2ik(y-x_2)}}{k}\, & x_1<x_2<y,\\[4mm]
\frac{e^{2ik(x_2-y)}+e^{2ik(y-x_1)}-2}{k}\, & x_1<y<x_2,\\[4mm]
\frac{e^{2ik(x_2-y)}-e^{2ik(x_1-y)}}{k}\, & y<x_1<x_2,
\end{cases}
$$
as $k\to \infty$. For large $R$, we may replace the integrand in the integral
over $\Gamma_R$ by these asymptotic values. Since there is
no singularity at $k=0$ we may deform the contour to an
integral over $(-R,R)$ on the real $k$-axis. The real parts,
involving cosines, are odd in $k$, and hence their contribution
vanishes. We are therefore left with the identity 
\begin{multline*}
 \lim_{R\to \infty}\frac{1}{2\pi i}\int_{\Gamma_R}
[\KK(x_2,y,k)-\KK(x_1,y,k)]\,8kdk \\[4mm]
=\lim_{R\to \infty} \int_{-R}^R
\frac{\sin 2k(y-x_1)-\sin 2k(y-x_2)}{2\pi k}dk\\[4mm]
=\frac12 [\text{sgn}(y-x_1)-\text{sgn}(y-x_2)].
\end{multline*}
 The proof of the lemma follows immediately when this result is 
substituted into the right side
of \reff{4}. \qed

The completeness theorem is proved by deforming the integral over
$\Gamma_R$  in \reff{4} to the real line. We begin by analyzing the poles of 
\beq{8kK}
8k\,\KK=-\vphi_+(x,k)\psi_+(y,k)\fr{R(x,y,k)}{ka^2(k)},
\qquad
x<y.
\end{equation}
Denote by $'$ differentiation with respect to $k$.

\blem{poles} The expression \reff{8kK} has a removable singularity at $k=0$ and simple
poles at $k=i\o_j$, $j=1,\dots,n$, with residues
\beq{res}
\KK_j(x,y)=-\fr{1}{2\o_j(a'(i\o_j))^2}
\big( F_j(x)G_j(y)-G_j(x)F_j(y)\big),
\end{equation}
for all $x,\,y$, where
\begin{multline}\label{FG}
F_j(x)=\psi_+^2(x,i\o_j),
\\[4mm]
G_j(y)=i\vphi_+(y,i\o_j)\fr{d}{dk}(\vphi_+(y,k)-c_j\psi_+(y,k))\Big|_{k=i\o_j},
\end{multline}
with $c_j$ such that $\vphi_+(x,i\o_j)=c_j\psi_+(x,i\o_j)$.
\elem

\Proof  
For $k=0$ the wave functions $\vphi_+$, $\psi_+$ are real so
$$
\vphi_-(x,0)=\vphi_+(x,0),\quad
\psi_-(x,0)=\psi_+(x,0).
$$
Then by \reff{scat} we obtain $R(x,y,0)=0$, hence \reff{8kK} is
regular
at $k=0$.

The poles of $8k\KK$ coincide with the zeros of $a$.
Hence, by \reff{ak}, it
has  $n$  poles, $i\o_j$, $j=1,\dots,n$.
 {From} the relation $\vphi_+(x,i\o_j)=c_j\psi_+(x,i\o_j)$,
where $c_j$ is the coupling coefficient defined by \reff{cm},
we see that $R(x,y,i\o_j)=0$ and $R'(x,y,i\o_j)\not=0$. 
Since $a^2(k)$ has double 
zeroes at $k=i\o_j$, $8k\,\KK$ has simple poles at $i\o_j$. A simple calculation shows
that the residues there are given by
\beq{resK}
\KK_j(x,y)=\begin{cases} -M_j(x,y),  & x<y \\ M_j(y,x), & y<x \end{cases}
\end{equation}
where
\beq{res2}
M_j(x,y)=\fr{\vphi_+(x,i\o_j)\psi_+(y,i\o_j)}{2i\o_j[a^\prime(i\o_j)]^2}B_j(x,y),
\qquad
x<y,
\end{equation}
and
\begin{multline}\label{Bj}
B_j(x,y)=\\[4mm]
\fr{d}{dk}\left(\vphi_+(x,k)\psi_+(y,k)-\psi_+(x,k)\vphi_+(y,k)\right)
\Big\vert_{k=i\o_j}
\quad x<y.
\end{multline}
 Using the relationship $\vphi_+(x,i\o_j)=c_j\psi_+(x,i\o_j)$, this form of $\KK_j$ can be rewritten in the form \reff{res}, 
as originally obtained by Sachs.\qed

We now deform the contour $\Gamma_R$ to the real axis, 
picking up the residues at $i\o_j$. By \reff{3}, the contribution
from the real line is
\begin{multline*}
\fr{1}{2\pi i}\inmp\inmp \KK(x,y,k)f(y)\,dy\,8k\,dk=\\[4mm]
\inmp\Big\{\frac{\vphi_+(x,k)\psi_+(x,k)}{ka^2(k)}
\Big[\inxp \vphi_+(y,k)\psi_+(y,k)f(y)\,dy
\\[4mm]
-\inmx \vphi_+(y,k)\psi_+(y,k) f(y)\,dy \Big]\\[4mm]
+ \frac{\psi_+^2(x,k)}{ka^2(k)} \inmx \vphi_+^2(y,k)f(y)\,dy
-\frac{\vphi_+^2(x,k)}{ka^2(k)} \inxp \psi_+^2(y,k)f(y)\,dy\Big\} 
\frac{dk}{2\pi i}.
\end{multline*}

From the equalities
$\psi_-(x,k)=\overline{\psi_+(x,k)}=\psi_+(x,-k)$
which hold for $k\in\R$,
by Sachs' argument \cite{Sa},  p. 678, this simplifies to
\beq{5}
\inmp \inmp (\psi_+^2(x,k)\psi_-^2(y,k)
-\psi_-^2(x,k)\psi_+^2(y,k))f(y)\,dy\,\fr{dk}{4\pi i k}.
\end{equation}
This gives the contribution from the real line; the contribution from the residues is
straightforward. Hence 

\btheo{c1} Let $u$ be an $n$-soliton potential.
Then for $f\in L_1(\mathbb R)$,
\begin{multline}\label{exp}
\int_{x_0}^{x}f(y)\,dy=
\sum_{j=1}^n\inmp (\KK_j(x_0,y)-\KK_j(x,y))f(y)\,dy \\[4mm]
\inmp\inmp\big[ (\psi_+^2(x,k)-
\psi_+^2(x_0,k))\psi_-^2(y,k)
\\[4mm]-(\psi_-^2(x,k)-\psi_-^2(x_0,k))
\psi_+^2(y,k)\big]f(y)\,dy\,\fr{dk}{4\pi i k} ,
\end{multline}
where $\KK_j$ are given in \reff{res}.
\end{theorem}
\btheo{c2} Let $f$ and its Fourier transform both be in
$L_1$. Then
\begin{align}
f(x)=&\inmp\inmp \left[D\psi_+^2(x,k)\psi_-^2(y,k)
-D\psi_-^2(x,k)\psi_+^2(y,k)\right]f(y)\,dy\,\fr{dk}{4\pi i k}\nonumber \\[4mm]
+&\sum_{j=1}^n C_j\inmp (DF_j(x)G_j(y)-DG_j(x)F_j(y))f(y)\,dy, \label{exp2}
\end{align}
where $F_j,\,G_j$ are given in \reff{FG} and 
\beq{Cj}
C_j=\fr1{2\o_j(a'(i\o_j))^2}.
\end{equation}
\end{theorem}

\Proof Since $\widetilde f(k)\in L_1$, $f$ is continuous everywhere, and
the derivative of the left side of \reff{exp} is $f(x)$ everywhere.
When $f$ and $\widetilde f$ are in $L_1$, the integrals on the right
side of \reff{exp2} converge absolutely. We may divide both sides of
\reff{exp} by $(x-x_0)$ and let $x\to x_0$, thereby obtaining \reff{exp2}. \qed

\section{Propagator of the linearized KdV equation}
\label{sec5}

Consider the homogeneous initial value problem
\beq{hom}
v_t+ v_{xxx}-6(uv)_x=0,
\qquad
v(x,s)=\phi(x). 
\end{equation}

The propagator for this time dependent equation is the operator
$T_{t,s}$ such that $v(t)=T_{t,s}\phi$.
By Theorem \ref{c2} the solution of the initial value problem 
\reff{hom} is
\begin{multline}\label{prop}
v(x,t)=\inmp \inmp \frac{1}{4\pi i k}
[ D\psi_+^2(x,t,k)\psi_-^2(y,s,k)-\\[4mm]
D\psi_-^2(x,t,k)\psi_+^2(y,s,k)] 
\phi(y)\,dy\,dk \\[4mm]
 + \sum_{j=1}^n C_j \inmp [DF_j(x,t)G_j(y,s)-
DG_j(x,t)F_j(y,s)]\phi(y)\,dy
=:T_{t,s}\phi,
\end{multline}
where 
\begin{gather*}
F_j(x,t)=\psi_+^2(x,t,i\o_j),\\[4mm]
G_j(y,s)=i\vphi_+(y,s,i\o_j)\fr{d}{dk}(\vphi_+(y,s,k)-c_j\psi_+(y,s,k))
\Big\vert_{k=i\o_j},
\end{gather*}
and $C_j$  is given by \reff{Cj}.

A direct computation shows that the  functions
$$
g_j(x,t)=\fr{d}{dk}(\vphi_+(x,t,k)-c_j\psi_+(x,t,k))\Big|_{k=i\o_j},
$$
satisfy both the Schr\"odinger equation and the equation $\psi_t=
B\psi$ (cf. \cite{Sa}). Then by Theorem \ref{l6}, $G_j$ satisfies 
the linearized potential KdV equation \reff{lpkdv} and 
$DG_j$ satisfies the linearized KdV equation
\reff{lkdv}.
The same is true for $F_j$. Hence the last term in \reff{prop} takes the form
of a projection 
onto the kernel of the linearized KdV equation.

The $n$-soliton solutions satisfy the KdV equation identically in the $2n$ parameters
$\a_1,\dots,\a_n$
and  $\o_1,\dots,\o_n$. Differentiating the KdV equation
with respect to each  of the $\a_1,\dots,\a_n$
and $\o_1,\dots,\o_n$ we obtain $2n$ solutions of the linearized KdV
equation \reff{lkdv}:
\beq{uao}
u_{\a_j}=\fr{\partial u}{\partial \a_j},\quad 
u_{\o_j}=\fr{\partial u}{\partial \o_j},\quad
j=1,\dots,n.
\end{equation}
It is easily seen that the $u_{\o_j}$ grow linearly with time.

We conjecture that the $2n$ functions $DF_j$ and $DG_j$ are linear combinations of \reff{uao}.
For the $DF_j$, from \reff{ruw}, we have 
$$
Du=-4\sum_{j=1}^n\o_je^{2\o_j\a_j}D(\psi_j^2)
=-4\sum_{j=1}^n\o_je^{2\o_j\a_j}DF_j,
$$
and by \reff{n-sol}
$$
Du=-\sum_{j=1}^n\fr{\partial
 u}{\partial \a_j}.$$
Hence, 
$$
\sum_{j=1}^n\o_je^{2\o_j\a_j}DF_j
=
\fr14\sum_{j=1}^n\fr{\partial
 u}{\partial \a_j}.
 $$
We conjecture that in fact 
$$
DF_j=
\fr{1}{4\o_j}e^{-2\o_j\a_j}\fr{\partial
 u}{\partial \a_j},
$$
holds for each $j=1,\dots,n$.
For $n=1,2$ we have confirmed this relationship by Maple calculations.

We expect each of the $DG_j$, $j=1,\dots,n$, to be a linear combination of 
all the functions \reff{uao}.
For the one-soliton solution we found, again by Maple calculations,
\beq{dg1}
8\o_1^2e^{-2\o_1\a_1}DG_1=
-\fr{\partial u}{\partial \o_1}+
\left(\fr{\a_1}{\o_1}-\fr{1}{2\o_1^2}\right)
\fr{\partial u}{\partial \a_1}.
\end{equation}
For the two-soliton solution the relationship between $DG_j$ and \reff{uao} seems already to be far more 
complicated, and we were unable to determine it with Maple.

We see from \reff{dg1} that 
$DG_1$ grows linearly in $t$,
since the derivative of the 
$n$-soliton solution with respect to $\o_1$
has this property.  In fact,  the same is true for all the $DG_j$ in the $n$-soliton case, as can be seen
by differentiating 
\reff{wf} with respect to $k$.
This implies that we obtain secular growth terms 
in the sum appearing in \reff{prop}. In order to 
eliminate  this secular growth in the propagator, we must impose 
the orthogonality
conditions 
\beq{ortc}
\inmp G_j(y,t)\phi(y)\, dy=
\inmp F_j(y,t)\phi(y)\,dy=0, \qquad 
j=1,\dots,n.
\end{equation}

Recall that $F_j$ and $G_j$ are solutions of the 
 the linearized potential KdV operator, which is the adjoint of the linearized KdV operator. 
So, the orthogonality conditions \reff{ortc} are 
analogous to solvability conditions in the Fredholm 
alternative for the time dependent operator occurring in \reff{hom}. (cf. Lemma \ref{pde} below.)

In the case of a single soliton, one may evaluate the linearized KdV equation in a frame
moving with the solitary wave. In that case the linearized equation is time independent, 
as are the pair of solutions $u_{\a_1}$ and $u_{\o_1}.$ This pair of  functions spans the
generalized
kernel of the linearized operator \cite{PW1}
$$
Lv:=v_{xxx}-4\o_1^2v_x-6(uv)_x.
$$
They are both exponentially decaying as $x\to\pm\infty$, hence are in $L_2(\mathbb R)$. 

In the present case, the linearized equation is no longer time independent, and we 
cannot talk about the kernel of a linear time independent operator. 
Nevertheless, the $2n$ functions \reff{uao} 
play a similar role in the analysis. They are exponentially decaying in $x$ for fixed time.

\section{Estimates on the  propagator}
\label{sec6}

In this section we estimate the propagator 
$T_{t,s}$ in spaces of functions analytic 
on the strip $|\Im x|<2\pi/\Omega$. 
Given a function $\phi$ analytic in this strip, define
$\phi_\a (x)=\phi(x+i\a)$, for $x\in \mathbb R$.
Consider the space
\begin{align*}
\AA_\a^m:=&\{ \phi \text{ analytic in } 
|\Im x|<\a,\ 
\phi_{\pm\a}
\in H^m(\R), \\[2mm]
&\quad \ \phi_y\to 0, \text{ uniformly in } |y|<\a, \text{ as }
|x|\to\infty
\}
\end{align*}
where  $0\leq\a<2\pi/\Omega$ and $m$ is some positive integer.
If $\a=0$ this space is  the Sobolev space $H^m(\mathbb R)$.
The norm in $\AA_\a^m$ is defined to be
$$
\|\phi\|_{\a,m}:=\max\{|\phi_\a |_m, |\phi_{-\a}|_m \} ,
$$
where $|\cdot|_m$ denotes the usual norm in $H^m(\mathbb R)$.

The following result follows from the properties of the Fourier
transform.

\blem{fou}
If $\phi\in \AA_\a^m$ then the Fourier transform $\wh \phi$ belongs
to 
$$
\wh \AA_\a^m:=\{\wh \phi(k)  : (1+|k|)^me^{\a|k|}
\wh \phi(k)\in L_2(\mathbb R)\}.
$$
Conversely, if $\wh \phi\in \wh \AA_\a^m$ 
then its inverse Fourier transform
$\phi$ belongs to $\AA_\a^m$.
Moreover, the norms of $\phi$ in $\AA_\a^m$ and of $\wh \phi$
in $\wh \AA_\a^m$ are equivalent.
\elem

The norm of $\wh \phi\in \wh \AA_\a^m$ is the $L_2$-norm of the
function $(1+|k|)^me^{\a|k|} 
\wh \phi(k)$, and we shall denote it also by $\|\cdot\|_{\a,m}$.

To analyze the propagator we write
$$
\psi_+^2=e^{2i(kx+4k^3t)}[1+w_+(x,t,k)],
\qquad
\psi_-^2=e^{- 2i(kx+4k^3t)}[1+w_-(x,t, k)].
$$
By the results in \S\ref{sec3}, 
$w_+$ is a meromorphic function of $k$, with poles in the
lower half plane at $-i\o_j,\  j=1,\dots\ n$, and it decays
like $1/k$ as $k$ tends to infinity. Moreover, $w_+$ is 
analytic  in $x$ in the strip $|\Im x|< 2\pi/\Omega$; it is
uniformly bounded in $|\Im x|\leq\a$, 
$t\in \mathbb R$,  $\Im k\geq -\o_1+\e$,
for any $0<\a<2\pi/\Omega$, $\e>0$,
and $w_-(x,t,k)=\overline{w_+(x,t,\bar k)}$. 

 Substituting the expressions for $\psi^2_{\pm}$ above
into the integrand in the first term of \reff{prop}, we find
\begin{align*}
\frac{1}{4\pi i k}&
[ D\psi_+^2(x,t,k)\psi_-^2(y,s,k)-D\psi_-^2(x,t,k)\psi_+^2(y,s,k)] \\[3mm]
=&\frac1\pi \left[\cos\,2(k(x-y)+4k^3 (t-s)) +
K_1(x,y,t,s,k)+K_2(x,y,t,s,k)\right],
\end{align*}
where
\begin{align*}
K_1(x,y,t,s,k)=&\fr{H_1(x,y,t,s,k)-H_1(y,x,s,t,k)}{2}, \\[4mm]
H_1(x,y,t,s,k)=&e^{2i(k(x-y)+4k^3 (t-s))}(w_+(x,t,k)\\[4mm]
& \hskip 30mm +w_-(y,s,k)+w_+(x,t,k)w_-(y,s,k)),\\[4mm]
K_2(x,y,t,s,k)=&
\fr{1}{4ik}[e^{2i(k(x-y)+4k^3 (t-s))}Dw_+(x,t,k)(1+w_-(y,s,k))\\[4mm]
&\hskip 10mm -e^{-2i(k(x-y)+4k^3 (t-s))}Dw_-(x,t,k)(1+w_+(y,s,k))].
\end{align*}
Since $\psi_+(x,t,0)=\psi_-(x,t,0)$, $w_+(x,t,0)=w_-(x,t,0)$,  
and $K_2$ has a removable singularity at $k=0$.

The propagator $T_{t,s}$ is thus a sum of 4 terms:
$$
T=T_0+T_1+T_2+T_3,
$$
where
\begin{align}
T_0(t,s)\phi=&\frac1\pi \inmp \inmp \cos\,2(k(x-y)+4k^3 (t-s))
\phi(y)\,dy\,dk ,\label{T0}
\\[4mm]
T_1(t,s)\phi=&\frac1\pi\inmp \inmp
K_1(x,y,t,s,k)) \phi(y)\,dy\,dk , \label{T1}
\end{align}
\begin{align}
T_2(t,s)\phi=&\frac1\pi \inmp\inmp 
K_2(x,y,t,s,k) \phi(y)\,dy\,dk ,\label{T2}\\[4mm]
T_3(t,s)\phi=& 
 \sum_{j=1}^n C_j \inmp
(DF_j(x,t)G_j(y,s)-\nonumber\\[4mm]
&\hskip 30mm DG_j(x,t)F_j(y,s))
\phi(y)\,dy. \label{T3}
\end{align}

We begin by estimating $T_0$. We first note that $T_0$ can be written
\begin{multline*}
\frac{1}{2\pi} \inmp \inmp e^{2i(k(x-y)+4k^3
  (t-s))}\phi(y)\,dy\,dk\\[4mm]
+\frac{1}{2\pi} \inmp \inmp e^{-2i(k(x-y)+4k^3 (t-s))}\phi(y)\,dy\,dk.
\end{multline*}
Changing $k \to -k$ in the second integral, and replacing
$2k$ by $k$  we get
\beq{T00}
T_0(t,s)\phi=\frac{1}{2\pi} \inmp  e^{i(kx+k^3 (t-s))}
\inmp e^{-iky}\phi(y)\,dy\,dk.
\end{equation}

The contribution in the solution of \reff{inh} coming  from $T_0$ is 
$$
v_0(x,t)=-\intp\fr{1}{2\pi}\inmp e^{ikx}e^{ik^3(t-s)}\inmp e^{-iky}
f(y,s)\,dy\,dk\,ds,
$$
or in terms of Fourier transforms
\beq{v0}
\widehat v_0(k,t)=-  \intp e^{ik^3 (t-s)}
\widehat f(k,s)\,ds.    
\end{equation}

The action of $T_0(t,s)$ on $\phi$ is multiplication 
of the Fourier transform
$\wh \phi$ by $e^{ik^3(t-s)}$.
This is a unitary operator on $\wh \AA_\a^m$, so by Lemma
\ref{fou} $T_0$ is a bounded operator  on $\AA_\a^m$, with bound 
independent of $t$ and $s$.

We next show that the operators $T_j(t,s),\ j=1,2$ are smoothing operators. 
 The first term in $T_1$ is in fact a pseudo-differential operator
$$
T_{1,1}(t,s)\phi= \frac1{2\pi}\inmp e^{ 2i(kx+4k^3(t-s))}
w_+(x,t,k) \widehat\phi(2k)\,dk.
$$
{From} the structure and decay properties in $k$ of $w_+$,  
$T_{1,1}$ is a bounded map from $\AA_\a^0$ to $\AA_\a^1$.
In fact, by \reff{wf}, $T_{1,1}$ is a sum of operators such as
$$
\fr1\pi\, p_j(x,t) \inmp e^{ 2i(kx+4k^3(t-s))} \frac{1}{k-i\o_j}
 \widehat\phi(2k)\,dk
$$
as well as others which are order $k^{-2}$. But multiplication of
$\widehat \phi$ by terms such as $(k-i\o)^{-1}$ is a smoothing
operator; it acts as an integration. Furthermore, multiplication by the
analytic  functions $p_j(x,t)$ 
is also a bounded operation on $\AA_\a^m$. 

We therefore see without difficulty  that
\beq{estT}
\|T_{1,1}(t,s)\phi\|_{\a,1}\le C\|\phi\|_{\a,0},
\end{equation}
for some constant $C$. Thus, by the smoothness of $w_\pm$ in $x$ and $y$ 
we see that $T_{1,1}$ maps continuously $\AA_\a^m$ to $\AA_\a^{m+1}$.
The other two terms can be treated in exactly
the same way. Actually, the third term in $T_1$ is a bounded map from
$\AA_\a^m$ to $\AA_\a^{m+2}$; though this fact is of no real use.
Similarly the operator $T_2$ is  a bounded map from $\AA_\a^m$ to $\AA_\a^{m+2}$,
since its kernel is regular at $k=0$ and decays like $k^{-2}$ as
$k\to \infty$.

The terms arising from $T_3$ grow linearly in time, as we observed in the previous section;
and so we have

\btheo{invkdv} Let $T_{t,s}$ be the propagator defined by the initial value problem
\reff{hom}, and let $\phi$ satisfy the orthogonality conditions \reff{ortc}. 
Then
$$
||T_{t,s}\phi||_{\a,m}\le C_{\a,m} ||\phi||_{\a,m}.
$$
\etheo

\section{Inversion of the linearized KdV equation}
\label{sec7}

We now turn to the inhomogeneous equation
\beq{inh}
v_t+ v_{xxx}-6(uv)_x=f(x,t),
\end{equation}
where $u$ is a multi-soliton solution of the KdV equation. 
By Duhamel's principle, solutions of this inhomogeneous equation
 are given
by
$$
v(\cdot,t)=-\int\limits_t^\infty T_{t,s}f(\cdot,s)\,ds .
$$
The solution of the inhomogeneous equation on the semi infinite interval 
$t>0$ is of course not unique. We have chosen this form so that
we preserve the class of functions decaying exponentially as
$t\to \infty$.

Let
$$
\RR^m_{\a,b}=\{R(x,t)  :  \sup_{t\geq 0}e^{bt}\|R(\cdot,t)\|_{\a,m}\leq\infty\},
$$
with  norm
$$
\|R\|_{\a,b,m}=\sup_{t\geq 0}e^{bt}\|R(\cdot,t)\|_{\a,m},
$$
where $\a$, $m$ are as in the previous section and  $b$ is a positive constant.
Denote by $\WW$ the class of all functions $f(x,t)$ of the form
\beq{wxt}
f(x,t)=\sum_{j=1}^nf_j(x-4\o_j^2t)+R_f(x,t),
\end{equation}
where $R_f\in \RR^m_{\a,b}$ and $f_j\in\AA^m_\a$. 
In fact we  assume that $f_j$ decays exponentially to zero as $|x|\to \infty$,
in the strip $|\Im x|\leq\a$, i.e. $f_j\in \AA^m_{\a,\mu}$, where
$$
\AA^m_{\a,\mu}=\{f  :  \, \cosh(\mu x)f\in \AA^m_\a\}.
$$
For simplicity we choose the same analyticity domain and the same 
exponential decay rate 
for all $f_j$, though this is not necessary.
In this section we solve the inhomogeneous linearized KdV equation in the 
space $\WW$.

The class $\WW$ corresponds, roughly, to a decomposition of the space of 
functions defined on the half plane $t>0$ into exponentially decaying functions
and persisting functions.

Recall (Theorem \ref{nasym}) that the $n$-soliton solution of the KdV equation 
is of the form 
\beq{uxt}
u(x,t)=\sum_{j=1}^nu_j(x-4\o_j^2t)+R_u(x,t),
\end{equation}
where $u_j(z)=-2\o_j^2\sech^2(\o_jz-\o_j\a_j+\gamma_j)$ and $R_u$ satisfies \reff{Rest}.
Hence, it belongs to $\WW$ for  any $0<\a<\pi/2\Omega$, $0\leq\mu<2\o_1$, for some
 $b>0$ as in \reff{Rest}, and 
any positive integer $m$.
 
We solve \reff{inh} in the space  $\WW$  above. 
For $f\in\WW$ we look for  solutions $v\in\WW$,
\beq{vxt}
v(x,t)=\sum_{j=1}^nv_j(x-4\o_j^2t)+V(x,t),
\end{equation}
where $V\in \RR^m_{\a,b}$ and $v_j\in \AA^m_{\a,\mu}$.
Since $u_j$ and $v_k$ 
decay exponentially in $x$, and  are waveforms moving to infinity at
different speeds, $u_jv_k\in\RR^m_{\a,b}$, for some $b>0$ related to the
difference $\o_j^2-\o_k^2$ of the  speeds of 
$u_j$ and $v_k$; thus waveforms moving at different
velocities decouple, and their product lies in $\RR^m_{\a,b}$ for some $b$.

The $v_j$ are determined independently by solving the ordinary differential equation
\beq{eqvj}
-4\o_j^2v_j^\prime+v_j^{\prime\prime\prime}-6(u_jv_j)^{\prime}
=f_j.
\end{equation}
Then for  $V$ we solve 
\beq{eqv}
V_t+V_{xxx}-6(uV)_x=G(x,t),
\end{equation}
where
\beq{g}
G(x,t)=6\sum_{j\neq k}(u_jv_k)_x+6\sum_{k=1}^n(R_u v_k)_x+R_f.
\end{equation}

\blem{ode}
Assume $f_j\in \AA^m_{\a,\mu}$, with $0<\a<\pi/2\Omega$,  
$0\leq\mu<2\o_1$, and 
\beq{cfj1}
\inmp f_j(z)\,dz=0.
\end{equation}
 Then, \reff{eqvj} has a unique solution 
$ v_j\in \AA_{\a,\mu}^{m+3}$ if and only if $f_j$ satisfies the orthogonality
condition
\beq{cfj}
\inmp u_j(z)f_j(z)\,dz=0.
\end{equation}
In particular, if  
$
f_j(z)=\widetilde f_j(\o_jz-\o_j\a_j+\gamma_j),
$
and  $\widetilde f_j$ is an even function in $\Re z$, 
then \reff{eqvj} has a unique solution
$v_j(z)=\widetilde v_j(\o_jz-\o_j\a_j+\gamma_j)$ with 
$\widetilde v_j$ an even function in $\Re z$.
\elem

\Proof
{From} \reff{cfj1} it follows that there exists $g_j\in \AA_{\a,\mu}^{m+1}$ such that
$f_j=g_j^\prime$.\footnote{This is precisely the situation which arises in our
perturbation scheme (unpublished here) of the Euler equations for surface gravity waves
which leads to the KdV approximation.}
Then by
integrating \reff{eqvj} once, we get
\beq{vj1}
v_j^{\prime\prime}-
4\o_j^2v_j-6u_jv_j=g_j.
\end{equation}

Recall that $u_j(z)=-2\o_j^2\sech^2(\o_jz-\o_j\a_j+\gamma_j)$. Then, set 
$y=\o_jz-\o_j\a_j+\gamma_j$ and \reff{vj1} reads
\beq{scaled}
 v_j^{\prime\prime}-4 v_j
+12\,\sech^2(y) v_j=h_j,
\end{equation}
with $h_j(y)=g_j(z)/\o_j^2$, $h_j\in \AA_{\a^\prime,\mu^\prime}^{m+1}$,
$\a^\prime=\o_j\a<\pi/2$, $\mu^\prime=\mu/\o_j<2$.

In the space above the operator $D^2+12\,\sech^2y-4$ is a compact perturbation of $D^2-4$,
since $0\leq\mu^\prime<2$. 
It is
easily seen  that $D^2-4$ possesses a bounded inverse from 
$\AA_{\a^\prime,\mu^\prime}^{m+1}$
into $\AA_{\a^\prime,\mu^\prime}^{m+3}$.
Hence the result also applies
to $D^2-4+12\,\sech^2y$ on the orthogonal complement of its kernel.

The operator $D^2-4+12\,\sech^2y$ is well-studied in connection with
the KdV solitons (cf. \cite{DJ}).
In $L_2(\mathbb R)$ its discrete spectrum consists of 
simple eigenvalues 5, 0, -3, and its continuous spectrum occupies
the interval $(-\infty,-4]$.
Due to translation invariance the derivative
of the solitary wave is a homogeneous solution of \reff{scaled}, i.e.,
$$
\phi_0=\sech^2x\ \tanh\, x,
$$
is a null function for \reff{scaled}. 
It spans the kernel of $D^2-4+12\,\sech^2y$ in $L_2(\mathbb R)$. 
But $\phi_0$ belongs to $ \AA_{\a^\prime,\mu^\prime}^{m+1}
\subset L_2(\mathbb R)$, so  
the kernel of $D^2-4+12\,\sech^2y$ in this space is also one dimensional.

{From} these considerations it follows that \reff{scaled} has a unique solution
provided $h_j$ is orthogonal to $\phi_0$. The same is true for the unscaled 
equation \reff{vj1}. A simple calculation shows that
the orthogonality  condition for \reff{vj1} is in fact \reff{cfj} and the first part of 
the theorem is proved. 

The final part  is a consequence of the fact that 
\reff{scaled} can be solved in spaces of even functions, and there  it
always has a unique solution. The eigenfunction  $\phi_0$ is  
odd, so the operator $D^2-4+12\,\sech^2y$ restricted to the subspace of even functions has a trivial kernel. \qed

We substitute now the $v_j$ obtained in this lemma into \reff{g}. 
Then $G\in \RR^m_{\a,b}$, where $b$ is less 
than the one given in \reff{Rest} and 
\beq{cb}
\min\{4\o_j(\o_{j+1}^2-\o_j^2),\  4\mu(\o_{j+1}^2-\o_j^2),\ 
j=1,\dots,n\}.
\end{equation}

\blem{pde}
Assume $0<\a<\pi/2\Omega$,  and  $G\in \RR^m_{\a,b}$ satisfies the
orthogonality conditions
\beq{ortcg}
\inmp G_j(y,t)G(y,t)\, dy=0, \qquad
\inmp F_j(y,t)G(y,t)\,dy=0, \qquad 
j=1,\dots,n,
\end{equation}
where $F_j$ and $G_j$ are the functions in \reff{FG}.
Then \reff{eqv} has a unique solution $V\in \RR^m_{\a,b}$, 
\beq{sv}
V(\cdot,t)=-\int_t^\infty (T_0+T_1+T_2)_{t,s}G(\cdot,s)\,ds,
\end{equation}
where $T_j$, $j=0,1,2$ are defined in \reff{T0}--\reff{T2}. 
\elem

\Proof
 Since $G$ satisfies \reff{ortcg}, 
we deduce from the results in \S\ref{sec5} that the solution $V$ of \reff{eqv}
is given by 
\reff{sv}. 
By the estimates in \S\ref{sec6} we have
$$
\|T_j(t,s)G(\cdot,s)\|_{\a,m}\leq C_j\|G(\cdot,s)\|_{\a,m},\quad
j=0,1,2,
$$
and the constants $C_j$ are independent of $t$ and $s$.
Then 
\begin{align*}
\|V\|_{\a,b,m}=&\sup_{t\geq 0}e^{bt}\int_t^\infty \|(T_0+T_1+T_2)_{t,s}
G(\cdot,s)\|_{\a,m}\,ds\\
\le& C\sup_{t\geq 0}e^{bt}\int_t^\infty 
e^{-bs}\,ds\,\|G\|_{\a,b,m}=\fr{C}{b}\|G\|_{\a,b,m},
\end{align*}
hence $V\in \R^m_{\a,b}$. \qed

The orthogonality conditions \reff{ortcg} are needed to eliminate terms which
grow linearly in time. Since $F_j$ and $G_j$ are solutions of the adjoint linearized
KdV equation, \reff{ortcg} may be viewed as Fredholm solvability conditions for
the inversion of the linearized KdV equation on a half space $t>0$.

From these two lemmas we have
\btheo{W}
Assume $u$ in \reff{n-sol} 
is an $n$-soliton solution of the KdV equation.
Then $u$ is of the form \reff{uxt}, 
and $u\in\WW$,  for any  
$0<\a<\pi/2\Omega$, $0\leq\mu<2\o_1$, 
 $b>0$ as in \reff{Rest}, and 
any positive integer $m$. 

Assume 
 $b$ is  less than  the value in \reff{cb}, and 
$f\in\WW$ is of the form \reff{wxt} where each $f_j$ satisfies
the orthogonality conditions \reff{cfj1} and \reff{cfj}. 
Then the inhomogeneous linearized KdV equation \reff{inh}
has a solution $v\in\WW$, given by \reff{vxt}, with $v_j$ solutions of 
\reff{eqvj},  provided that
$$
G(x,t)=6\sum_{j\neq k}(u_jv_k)_x+6\sum_{k=1}^n(Rv_k)_x+R_f,
$$
satisfies \reff{ortcg}.
\etheo

In their analysis of the stability of the solitary wave for 
the generalized KdV equation, Pego and Weinstein \cite{PW1}
decomposed the equations for the perturbation into the 
domain and range of the linearized operator, and 
obtained the orthogonality conditions by imposing modulation equations
on the phase $\a$ and speed $\o$ of the solitary wave. 
Such a perturbation scheme is in principle feasible in 
the case of the multi-solitons; but the fact that the 
perturbation scheme loses derivatives make the problem
considerably more complicated, as we indicated in the introduction.

\section{Estimates in weighted spaces}
\label{sec8}

In \cite{PW1} weighted spaces have been considered to prove the asymptotic
stability of the solitary wave solutions of the KdV equation.
In these spaces the inverse of the linearized KdV operator for solitary waves
was a smoothing operator.
In this section we  consider similar weighted spaces but for  functions
which are analytic in a strip. We estimate the propagator $T$ and then
  the solution of \reff{inh} in these spaces.

Define
the weighted space
\beq{asAa}
\BB_{\a,\eta}^m:=\{ \phi   :  e^{\eta x}
\phi \in \AA^m_\a(\mathbb R)\},
\end{equation}
with norm
$
\|\phi\|_{\a,\eta,m}:=
\|e^{\eta x}\phi\|_{\a,m}.
$
Consider also
\beq{asLa}
\BB^m_{\a,\eta,b}:=\{f(x,t)  :  \sup_{t\geq 0}e^{b t}
\|f(\cdot,t)\|_{\a,\eta,m}<\infty \}.
\end{equation}
Denote by $\|\cdot\|_{\a,\eta,m,b}$ the norm in
this space.

{From} the properties of the Fourier
transform one obtains the following result.

\blem{fou2}
If $\phi\in \BB_{\a,\eta}^m$ then its Fourier transform belongs to
$
\wh \BB_{\a,\eta}^m:=\{\wh \phi(k)   : 
(1+|k|)^m
e^{\a|k|}\wh \phi(k+i\eta)\in L_2(\mathbb R)\}.$
 The spaces $\wh \BB_{\a,\eta}^m$
and $\BB^m_{\a,\eta}$ are isomorphic by the Plancherel theorem.
\elem

The norm in $\wh \BB_{\a,\eta}^m$, being equivalent with the one in 
$\BB_{\a,\eta}^m$,
 is also denoted by 
$\|\cdot\|_{\a,\eta,m}$.

To obtain estimates on the propagator $T$ in these spaces we have to 
consider  the KdV equation in a moving frame. 
We shall take  the following 
form of the  KdV equation
\beq{kdv1}
u_t-\fr16 u_{xxx}+\fr12 u_x+\fr32 uu_x=0.
\end{equation}
This is the actual equation we obtained in the long wave approximation of the Euler equations for water waves

The inhomogeneous linearized 
equation we solve is 
\beq{inh1}
v_t-\frac16 v_{xxx}+\frac12 v_x+\frac32 (uv)_x=f(x,t),
\end{equation}
with $u$  an $n$-soliton solution of \reff{kdv1}. 
The results in the previous sections are all valid for \reff{inh1}.

The solution of \reff{inh1} is given by 
$$
v(\cdot,t)=-\int\limits_t^\infty T_{t,s}f(\cdot,s)\,ds ,
$$
where $T$ is the propagator  of the homogeneous equation. It is defined as
for \reff{inh} but with different  wave functions 
$\psi_+$ and $\vphi_+$, 
$$
\vphi_+(x,t,k)=e^{- i(kx-\sigma t)}(1+m(x,t,k)),
\quad
\psi_+(x,t,k)=e^{i(kx-\sigma t)}(1+r(x,t,k)),
$$
where 
$$
\sigma =\frac23 k^3+\frac12 k.
$$
The reduced wave functions $m$ and $r$ are normalized by \reff{norm}.
They have the same properties as the reduced wave functions in 
\S\ref{sec3}.

The propagator $T_0$ is 
\begin{align}
T_0(t,s)\phi=&{\frac1\pi \inmp \inmp \cos\,2(k(x-y)-\sigma (t-s))
\phi(y)\,dy\,dk}\label{t0}\\[4mm]
=&{\frac{1}{2\pi} \inmp  e^{i(kx+\theta (s-t))}
\inmp e^{-iky}\phi(y)\,dy\,dk,}
\end{align}
where $\sigma=\frac{ k^3}{6}+\frac{k}{2}$.
The other terms in $T$, $T_1$, $T_2$ and $T_3$, have the same structure as before.

Choose $\a$ such that the decomposition 
\reff{ms}  and the estimate \reff{Rest} hold in the strip
$|\Im x|\leq \a$.

\btheo{t91}
Assume $0<\eta <\sqrt{3}$, $b>-\frac16\eta(3-\eta^2)$, and $m\geq 1$ is an integer.
Then the propagator $T_0$
satisfies:
\beq{97}
\left\|\intp T_0(t,s)f(\cdot,s)\,ds \right\|_{\a,\eta,m,b}
\leq C_1(\eta,b)
\| f\|_{\a,\eta,m-1,b},
\end{equation}
where $C_1(\eta,b)$ is a positive constant,
and $C_1(\eta,b)\rightarrow\infty$, as
$\eta\rightarrow 0$ and $\eta\rightarrow\sqrt 3$.
\etheo

\Proof
The propagator $T_0$ from \reff{t0} is the multiplication of the Fourier 
transform $\wh\phi$ by $e^{i\s (s-t)}$. Then, to obtain \reff{97} 
we first  need  bounds for $e^{i\s (s-t)}$ and  
$ke^{i\s (s-t)}$, for $\Im k=\eta$.

Let $k=\xi+i\eta$, $0<\eta<\sqrt{3}$, and $s>0$. We have
$$
| e^{i\s s}|=e^{-\Im\s s}=
e^{-\frac16\eta(3(\xi^2+1)-\eta^2)s}
\leq
e^{-\frac16\eta(3-\eta^2)s},
$$
and
$$
| ke^{i\s s}|=\sqrt{\xi^2+\eta^2}
e^{-\frac16\eta(3(\xi^2+1)-\eta^2)s}
\leq c_0(\eta,s),
$$
where
$$
c_0(\eta,s)=
\begin{cases}\fr{1}{\sqrt{e\eta s}}
e^{-\frac16\eta(3-4\eta^2)s}, &
0<s<\fr{1}{\eta^3}
\\[2mm]
\eta e^{-\frac16\eta(3-\eta^2)s}, &
s\ge \fr{1}{\eta^3}.
\end{cases}
$$

Hence, we find, for $s\geq t$,
$$
\|\wh{T_0(t,s)f(s)}\|_{\a,\eta,m}\leq c_0(\eta,s-t)c_1
\|\wh f(s)\|_{\a,\eta,m-1},
$$
for some $c_1>0$, 
and by Lemma \ref{fou2} 
$$
\|{T_0(t,s)}f(s)\|_{\a,\eta,m}\leq c_0(\eta,s-t)c_2
\|f(s)\|_{\a,\eta,m-1}.
$$

Denote
$$
v_0(x,t)=\intp T_0(t,s) f(s)\,ds.
$$
Then
\begin{align*}
e^{b t}\| v_0(t)\|_{\a,\eta,m}\le & c_2\intp e^{bt}c_0(\eta,s-t)
\|f(s)\|_{\a,\eta,m-1}\,ds \\[4mm]
\le & 
c_2\intp e^{b(t-s)}c_0(\eta,s-t)\,ds \|f\|_{\a,\eta,m-1,b}\\[4mm]
= &
c_2 \int_0^\infty e^{-b s}c_0(\eta,s)\,ds \,\|f\|_{\a,\eta,m-1,b}
 = C_1(\eta,b)\|f\|_{\a,\eta,m-1,b} .
\end{align*}
It is easily seen that $C_1(\eta,b)<\infty$, for $0<\eta<\sqrt3$, and $b$ 
as in the theorem. Furthermore,
 $C_1(\eta,b)\to \infty$ as $\eta \to 0,\,\sqrt3$.
\qed

The operators $T_j,\ j=1,2$ are again smoothing operators.
In fact by arguing as in \S\ref{sec6} we can prove
\beq{97an}
\|\intp T_j(t,s)f(s)\,ds \|_{\a,\eta,m,b}
\leq C_1(\a,\eta,b)
\| f\|_{\a,\eta,m-j-1,b}.
\end{equation}

{From} these results follows.

\btheo{ws}
Assume $\eta$ and $b$ are as in Theorem \ref{t91}. If
$f\in \BB_{\a,\eta,b}^m$ satisfies the orthogonality conditions 
\reff{ortc}, then  \reff{inh1} has a  solution 
$v\in \BB_{\a,\eta,b}^{m+1}$,
$$
v(t)=-\sum_{j=0}^2\intp T_j(t,s)f(s)\,ds,
$$
and
$$
\|v\|_{\a,\eta,b,m+1}\leq C\|f\|_{\a,\eta,b,m}.
$$
\etheo

This theorem is the analog of Lemma \ref{pde} for the weighted spaces. 
A result similar to the one in Theorem \ref{W} holds then 
for a space $\WW_\eta$, defined as $\WW$ but with $R_f\in \BB_{\a,\eta,b}^m$,
instead of $R_f\in \AA_{\a,b}^m$.


\newpage
 

\begin{thebibliography}{10}

\bibitem{AK}
{\sc Amick, C.J. \& Kirchg\"assner, K.}
\newblock A theory of solitary water-waves in the presence of surface tension.
\newblock {\em {\it Arch. Rat. Mech. Anal.}}, {\bf 105}:1--49, (1989).

\bibitem{Be}
{\sc Beale, J.T.}
\newblock The existence of solitary water waves.
\newblock {\em {\it Comm. Pure Appl. Math.}}, {\bf 30}:373--389, (1977).

\bibitem{BL}
{\sc Bona, J.L. \& Li, Yi}.
\newblock Decay and analyticity of solitary waves.
\newblock to appear, {\it J. Math. Pures et Appl.}

\bibitem{Bo}
{\sc Boussinesq, M.J.}
\newblock Essai sur la th\'eorie des eaux courantes.
\newblock {\em {\it M\'emoires pr\'esent\'es par divers savants \`a
  l'Acad\'emie des Sciences Inst. France (s\'eries 2)}}, {\bf 23}:1--680,
  (1877).

\bibitem{Cr}
{\sc Craig, W.}
\newblock An existence theory for water waves and the boussinesq and
  korteweg-devries scaling limits.
\newblock {\em {\it Commun. Partial Diff. Eq.}}, {\bf 10}:787--1003, 1985.

\bibitem{GGKM}
{\sc Gardner, C.S., Green, J.M., Kruskal, M.D., \& Miura, R.M.}
\newblock Korteweg-devries equation and generalizations. vi. methods for exact
  solution.
\newblock {\em {\it Comm. Pure. Appl. Math.}}, {\bf 27}:97--133, (1974).

\bibitem{KN}
{\sc Kano, T., \& Nishida, T.}
\newblock A mathematical justification for korteweg-de vries equation and
  boussinesq equation of water surface waves.
\newblock {\em {\it Osaka J. Math.}}, {\bf 23}:389--413, (1986).

\bibitem{Ka}
{\sc Kato, T.}
\newblock On the cauchy problem for the (generalized) korteweg-de vries
  equation.
\newblock {\em {\it Stud. in Appl. Math. Suppl. Ser.}}, {\bf 8}:93--128,
  (1983).

\bibitem{Ki}
{\sc Kirchg\"{a}ssner, K.}
\newblock Nonlinearly resonant surface waves and homoclinic bifurcation.
\newblock {\em {\it Adv. Appl. Mech.}}, {\bf 26}:135--181, (1988).

\bibitem{KdV}
{\sc Korteweg, D.J. \& G. de Vries, G.}
\newblock On the change of form of long waves advancing in a rectangular canal,
  and on a new type of long stationary waves.
\newblock {\em {\it Phil. Mag.}}, {\bf 39}:422--443, (1895).

\bibitem{madsachs}
{\sc Maddocks, J. and R.L. Sachs}.
\newblock On the stability of kdv multi-solitons.
\newblock {\em Comm. Pure Appl. Math.}, 46:867--901, 1993.

\bibitem{Mo}
{\sc Moser, J.}
\newblock {A rapidly convergent iteration method and nonlinear partial
  differential equations - I,II.}
\newblock {\em {\it Ann. Scuola Norm. Sup. Pisa}}, {\bf 20}:265--315, 499--535,
  (1966).

\bibitem{PW1}
{\sc Pego, R.L. \& Weinstein, M.I.}
\newblock Asymptotic stability of solitary waves.
\newblock {\em {\it Comm. Math. Phys.}}, {\bf 164}:305--349, (1994).

\bibitem{PW2}
{\sc Pego, R.L. \& Weinstein, M.I.}
\newblock Convective linear stability of solitary waves for boussinesq
  equations.
\newblock preprint, 1996.

\bibitem{DJ}
{\sc P.G. Drazin, P.G. \& Johnson, R.S.}
\newblock {\em Solitons: an introduction}.
\newblock Cambridge University Press, Cambridge, (1989).

\bibitem{Ru}
{\sc Russell, J. Scott }.
\newblock Report on waves.
\newblock In John Murray, editor, {\em {\it Rept. Fourteenth Meeting of the
  British Association for the Advancement of Science}}, pages 311--390, London,
  1844.

\bibitem{Sa}
{\sc Sachs, R.L.}
\newblock {Completeness of derivatives of squared Schr\"odinger eigenfunctions
  and explicit solutions of the linearized KdV equation}.
\newblock {\em {\it SIAM J. Math. Anal.}}, {\bf 14}:674--683, (1983).

\bibitem{Sc}
{\sc Scheurle, J.}
\newblock Newton iterations without inverting the derivative.
\newblock {\em {\it Math. Meth. in the Appl. Sci.}}, {\bf 1}:514--529, (1979).

\end{thebibliography}
\end{document}